\documentclass[prd,twocolumn,showpacs,superscriptaddress,eqsecnum,floatfix,nofootinbib]{revtex4}

\usepackage{latexsym}
\usepackage{amssymb}
\usepackage{amsfonts}
\usepackage{amsmath}
\usepackage[dvips]{graphicx}

\newcommand{\be}{\begin{equation}}  
\newcommand{\ee}{\end{equation}}
\newcommand{\bea}{\begin{eqnarray}}           
\newcommand{\eea}{\end{eqnarray}} 
\newcommand{\ba}{\begin{align}}
\newcommand{\ea}{\end{align}}
\newcommand{\beqn}{\begin{eqnarray*}}
\newcommand{\eeqn}{\end{eqnarray*}}
\def\de{\partial}

\begin{document}
  
  \title{Dynamical excitation of space-time modes of compact objects} 
  
  \author{Sebastiano \surname{Bernuzzi}}
  \affiliation{Dipartimento di Fisica, Universit\`a di Parma, Via
    G.~Usberti 7/A, 43100 Parma, Italy} 
  \affiliation{INFN, Gruppo
    Collegato di Parma, Italy} 
  \author{Alessandro \surname{Nagar}}
  \affiliation{Institut des Hautes Etudes Scientifiques, 91440
    Bures-sur-Yvette, France} 
  \affiliation{INFN, Sezione di Torino,
    Via P.~Giuria 1, Torino, Italy} 
  \author{Roberto \surname{De Pietri}} 
  \affiliation{Dipartimento di Fisica, Universit\`a di
    Parma, Via G.~Usberti 7/A, 43100 Parma, Italy} 
  \affiliation{INFN, Gruppo Collegato di Parma, Italy}
  
  \date{\today}
  \begin{abstract}
    We discuss, in the perturbative regime, the scattering of Gaussian
    pulses of odd-parity gravitational radiation off a non-rotating
    relativistic star and a Schwarzschild Black Hole.  We focus on the
    excitation of the $w$-modes of the star as a function of the width
    $b$ of the pulse and we contrast it with the outcome of a
    Schwarzschild Black Hole of the same mass.  For sufficiently narrow
    values of $b$, the waveforms are dominated by characteristic
    space-time modes.  On the other hand, for sufficiently large values of
    $b$ the backscattered signal is dominated by the tail of the
    Regge-Wheeler potential, the quasi-normal modes are not excited and
    the nature of the central object cannot be established.  We view this
    work as a useful contribution to the comparison between perturbative
    results and forthcoming $w$-mode 3D-nonlinear numerical simulation.
  \end{abstract}
  
  \maketitle
  
  \section{Introduction}
  \label{intro}
  
  The pioneering works of Vishveshwara~\cite{vishveshwara70},
  Press~\cite{press72} and Davis, Ruffini and Tiomno~\cite{DRT}, 
  unambiguously showed that a non-spherical gravitational perturbation of
  a Schwarzschild Black Hole is radiated away via exponentially damped
  harmonic oscillations. These oscillations are interpreted as space-time
  vibrational modes. The properties of these {\it quasi-normal modes}
  (QNMs henceforth) of Black Holes have been thoroughly studied since
  then (see for example Refs.~\cite{chandra,fn_98,ks_99} and references
  therein). Relativistic stars can also have space-time vibrational
  modes, the so-called $w$-modes~\cite{ks_92}. These modes are purely
  relativistic and, contrary to fluid modes, are absent in Newtonian
  theory. The fundamental $w$-mode frequency of a typical neutron star 
  of radius~$\sim 10$ km and mass $1.4M_{\odot}$ is expected to lie
  in the range of $10\div12$ kHz and to have a damping time 
  of~$\sim10^{-4}$ s~\cite{Andersson:2004bi}.
  
  The issue of the excitation of $w$-modes in astrophysically motivated
  scenarios has been deeply investigated in the literature. 
  Andersson and Kokkotas~\cite{ak_96} showed that, in the odd-parity case, 
  the scattering of a Gaussian pulse of gravitational waves off a constant
  density non rotating star generates a waveform that, in close analogy
  with the Black Hole case, is characterized by three phases: (i) a
  precursor, mainly related to the choice of the initial data and
  determined by the backscattering of the background curvature while
  the pulse is entering in the gravitational field of the star; (ii) a burst; 
  (iii) a ring-down phase dominated by $w$-modes, whose presence was inferred 
  by looking at the Fourier spectrum of the signals.  
  Since the star is non-rotating, the signal eventually dies
  out with a power-law tail typical of Schwarzschild 
  space-time~~\cite{price1972,price1972bis}.
  Allen and coworkers~\cite{allen98} and Ruoff~\cite{ruoff_01a} 
  addressed, by means of time-domain perturbative analysis, 
  the same problem in the even-parity case, focusing on 
  gravitational wave scattering scenarios.
  They considered a large sample of initial configurations 
  as well as star models of different compaction. 
  Their main findings were: (i) 
  $w$-modes are present only for non-conformally flat initial data 
  (i.e., some radiative field needs to be injected in the system) 
  and (ii) the strength of the $w$-mode signal depends on 
  the compaction of the star. These pioneering studies were later 
  extended or refined in Refs.~\cite{Tominaga:1999iy,fk_00,Tominaga:2001,ruoff_01b,passamonti05a,
    passamonti06a,Pons:2001xs,Gualtieri:2001cm,Benhar:1998au,andersson97,andersson95b}.
  In particular, Refs.~\cite{Tominaga:1999iy,fk_00,Tominaga:2001,ruoff_01b} 
  considered the scattering off the star of particles moving along open orbits 
  and realized that the $w$-mode excitation strongly depends on the orbital parameters: 
  the closer the turning point of the orbit is to the star 
  (i.e., the higher is the frequency of the gravitational 
  wave instantaneously emitted by the particle), 
  the larger is the presence of $w$-modes.
  Consistently, Ref.~\cite{nagar04a} showed that (modulo a
  simplified treatment of the star surface) if the source of 
  perturbation is a spatially extended axisymmetric distribution 
  of fluid matter (like a quadrupolar shell) plunging on the star, 
  the $w$-modes are not excited, but the energy spectrum is dominated 
  by low-frequency contributions due to curvature backscattering.
  In addition, Ref.~\cite{pavlidou_00} addressed the late-time 
  decay of the {\it trapped} mode  for ultra-compact, highly relativistic 
  constant density stars. The presence of trapped $w$-modes in stars with a
  first-order phase transition (a density discontinuity) was also
  discussed in Ref.~\cite{Andrade:2001hk}.
  
  In this work we analyze the problem of $w$-modes excitation 
  in relativistic stars (in the perturbative regime) 
  by emphasizing the analogies with the Black Hole case.
  For a given odd-parity gravitational wave multipole $\ell$, we 
  consider the scattering of Gaussian pulses of gravitational radiation 
  of different width $b$ off relativistic stars 
  (either with constant density or with a polytropic equation of state) 
  of 1.4 $M_{\odot}$ and we contrast such signals with those emitted 
  by a non-rotating Black Hole of the same mass. 
  We focus on the excitation of space-time modes as a function of the 
  width $b$ of the Gaussian. We find that space-times modes can 
  be clearly identified only if the Gaussian wave-packet 
  is sufficiently narrow (i.e., small $b$). On the other hand, for 
  large wave-packets the internal structure of the object is 
  unaffected by the perturbation, tail effects are dominating and 
  the gravitational waveforms generated by stars or Black Holes 
  are practically identical.
  
  \section{Numerical framework}
  \label{sec:analytic}
  
  \subsection{Relativistic stars}
  \label{sbsc:bckg}
  
  From the spherically symmetric line element in Schwarzschild coordinates 
  \begin{equation}
    ds^2 = -e^{2\alpha}dt^2 + e^{2\beta}dr^2 + r^2\left(d\theta^2 + \sin^2\theta d\varphi^2\right) ,
  \end{equation}
  by assuming the stress energy tensor of a perfect fluid as
  $T^{\mu\nu} = (p+\mu)u^\mu u^\nu + p g^{\mu\nu}$,
  where $p$ is the pressure and $\mu$ the total energy density
  of the star, the Einstein equations reduce to the 
  Tolman-Oppenheimer-Volkoff (TOV) equations of stellar equilibrium:
  \begin{align}
    \label{tov1}
    \dfrac{dm}{dr} & = 4\pi r^2 \mu \ , \\
    \label{tov2}
    \dfrac{da}{dr} & = \frac{(m+4\pi r^3 p)}{(r^2 -2mr)}\ ,\\
    \label{tov3}
    \dfrac{dp}{dr} & = -(p+\mu)\frac{d\alpha}{dr}\ .
  \end{align}
  Since we are using Schwarzschild coordinates, we also have
  that $e^{-2\beta} = 1-2m(r)/r$, where $m(r)$ is the mass contained 
  in a sphere of radius $r$. This system of equations needs, to be solved, 
  the specification of an Equation of State (EoS). 
  We use the simplest two: the constant density EoS and an adiabatic 
  EoS in the form $p=K\mu^{\Gamma}$. We consider two polytropic  models 
  (named A and B) and two constant energy density models (A$_{\rm C}$ and
  B$_{\rm C}$) with the same compactness. All the models share the same 
  mass  $M=1.4M_\odot$, and their  specific properties are listed in 
  Table~\ref{tab:table1}. If not differently stated, we use geometrized 
  units $c=G=1$ with $M_{\odot}=1$.
  
  \begin{table}[t]
    \caption{\label{tab:table1} From up to down the rows report: the polytropic constant 
      $K$, the adiabatic index $\Gamma$, the mass of the star $M$, its radius $R$, the 
      central pressure $p_{\rm c}$, the central total energy density $\mu_{\rm c}$ and the 
      compaction parameter $M/R$, for all the stellar models considered.}
    \begin{ruledtabular}
      \begin{tabular}{ccccc}
        EoS  &   A & B & A$_{C}$ & B$_{C}$ \\
        \hline
        \hline
        $K$            &  $56.16$               &             $82.69$       &       ---                 &    ---                  \\
        $\Gamma$       &  $2$                  &             $ 2$          &       ---                 &    ---                  \\
        $M$            &   $1.40$              &             $ 1.40$       &     $1.40$                  &   $ 1.40$                 \\
        $R$            &   $6.64$              &             $ 9.10$       &     $ 6.64$                 &   $ 9.10$                 \\
        $p_{\rm c}$    &  $8.84\times 10^{-4}$ &   $1.83\times 10^{-4}$  &     $2.13\times 10^{-4}$  &   $4.98\times 10^{-5}$  \\
        $\mu_{\rm c}$  &  $3.96\times 10^{-3}$ &   $1.48\times 10^{-3}$  &     $1.14\times 10^{-3}$  &   $4.43\times 10^{-4}$  \\
        $M/R$          &    $  0.21  $        &     $   0.15 $            &     $0.21$                  &    $ 0.15 $               \\
      \end{tabular}
    \end{ruledtabular}
  \end{table}
  
  \subsection{Odd-parity perturbations}
  \label{sbsc:perturbs}
  
  Odd-parity linear perturbations of Black Holes and neutron stars 
  (in the absence of external matter source) are described by a simple linear equation 
  \begin{equation}
    \label{eq:wave}
    \de^2_{t}\Psi^{(\rm o)}-\de^2_{r_*}\Psi^{(\rm o)} + V^{(\rm o)}_{\ell}\Psi^{(\rm o)} = 0 \ ,
  \end{equation}
  for a master function $\Psi^{(\rm o)}\equiv\Psi^{(\rm o)}_\ell$
  that is related to the metric degrees of freedom (see for example
  Ref.~\cite{NR05}). For a star of radius $R$, the potential is given by
  \begin{equation}
    V^{(\rm o)}_{\ell} = e^{2a} \left( \frac{6m}{r^3} + 4\pi(p -\mu)  - \cfrac{\ell(\ell+1)}{r^2} \right) \ ,
  \end{equation}
  which reduces to the standard Regge-Wheeler potential for $r>R$, where
  $M=m(R)$ is the total mass and $p=\mu=0$. The latter holds also for the Black Hole
  case. We expressed Eq.~(\ref{eq:wave}) using a $r_*$ tortoise coordinate 
  defined as $d r_*/d r = \exp(\beta-\alpha)$. This reduces in vacuum to the 
  Regge-Wheeler tortoise coordinate $r_* = r + 2M \ln(r/(2M) -1)$. 
  The power emitted in gravitational waves is: 
  \begin{equation}
    \label{eq:energy}
    \dot{E}^{(\rm o)} \equiv \sum_{\ell\geq 2} \dot{E}^{(\rm o)}_{\ell}=\sum_{\ell\geq 2}\dfrac{(\ell+2)!}{(\ell-2)!}
    |\dot{\Psi}^{(\rm o)}_{\ell}|^2 \ ,
  \end{equation}
  where the over-dot stands for coordinate time derivative.
  
  \subsection{Initial data and simulation method}
  \label{sec:initialdata}
  
  For NS and Black Holes Eq.~(\ref{eq:wave}) is solved in the time
  domain as an initial value problem. In the case of the polytropic 
  EOS, we need first to integrate numerically the TOV equations 
  (\ref{tov1})-(\ref{tov3}) to compute the potential $V^{(\rm o)}_{\ell}$. 
  For a given central pressure $p_{\rm c}$ (see Table~\ref{tab:table1}) 
  the TOV equations are integrated numerically (from the center outward)
  using a standard fourth-order Runge-Kutta integration scheme with 
  adaptive step size. 
  
  As initial data for ($\Psi^{(\rm o)},\de_t\Psi^{(\rm o)}$)
  we set up an ingoing ($\de_t\Psi^{(\rm o)}=\de_{r_*}\Psi^{(\rm o)}$) 
  Gaussian pulse of tunable width $b$
  \begin{equation}
    \label{eq:ID}
    \Psi^{(\rm o)}={\cal N}\exp\left[-(r-r_0)^2/b^2\right] \ ,
  \end{equation}
  where ${\cal N}$ is a normalization constant determined by equating 
  to one the integral of Eq.~(\ref{eq:ID}) all over the radial domain.
  This is a simple, but sufficiently general, way to represent a 
  ``distortion'' of the space-time (whose intimate origin depends on 
  the particular astrophysical setting), and to introduce in the system 
  a proper scale through the width of the Gaussian.
  
  Let us also summarize the basilar elements of our numerical procedure.
  Eq.~(\ref{eq:wave}) has been discretized on an evenly spaced grid 
  (in $r_*$ for the Black Hole and in $r$ for the star) and solved using 
  a standard implementation of the second-order Lax-Wendroff method as
  implemented for example in~\cite{nagar04a}. We have performed convergence
  tests of the code which assured a  convergence factor of $\sim 2$.
  A resolution of $\Delta r_*=0.01$ and $\Delta r=0.015$ is sufficient to 
  be in the convergence regime. Since we have implemented standard 
  Sommerfeld outgoing boundary conditions (see Ref.~\cite{gr-qc/0507140} for
  improved, non-reflecting boundary conditions), we can't avoid some 
  spurious reflections to come back from boundaries. To avoid that this 
  effect contaminates too much the late-time tails of the signals, 
  we need to choose radial grids sufficiently extended, 
  say  $r_*\in[-2000,2000]$ and $r\in[0,2000]$. 
  
  \section{Results}
  \label{sec:results}
  
  \subsection{Analysis of the waveforms}
  \label{sbsc:waveforms}
  
  \begin{figure}
    \begin{center}
      \includegraphics[width=85 mm]{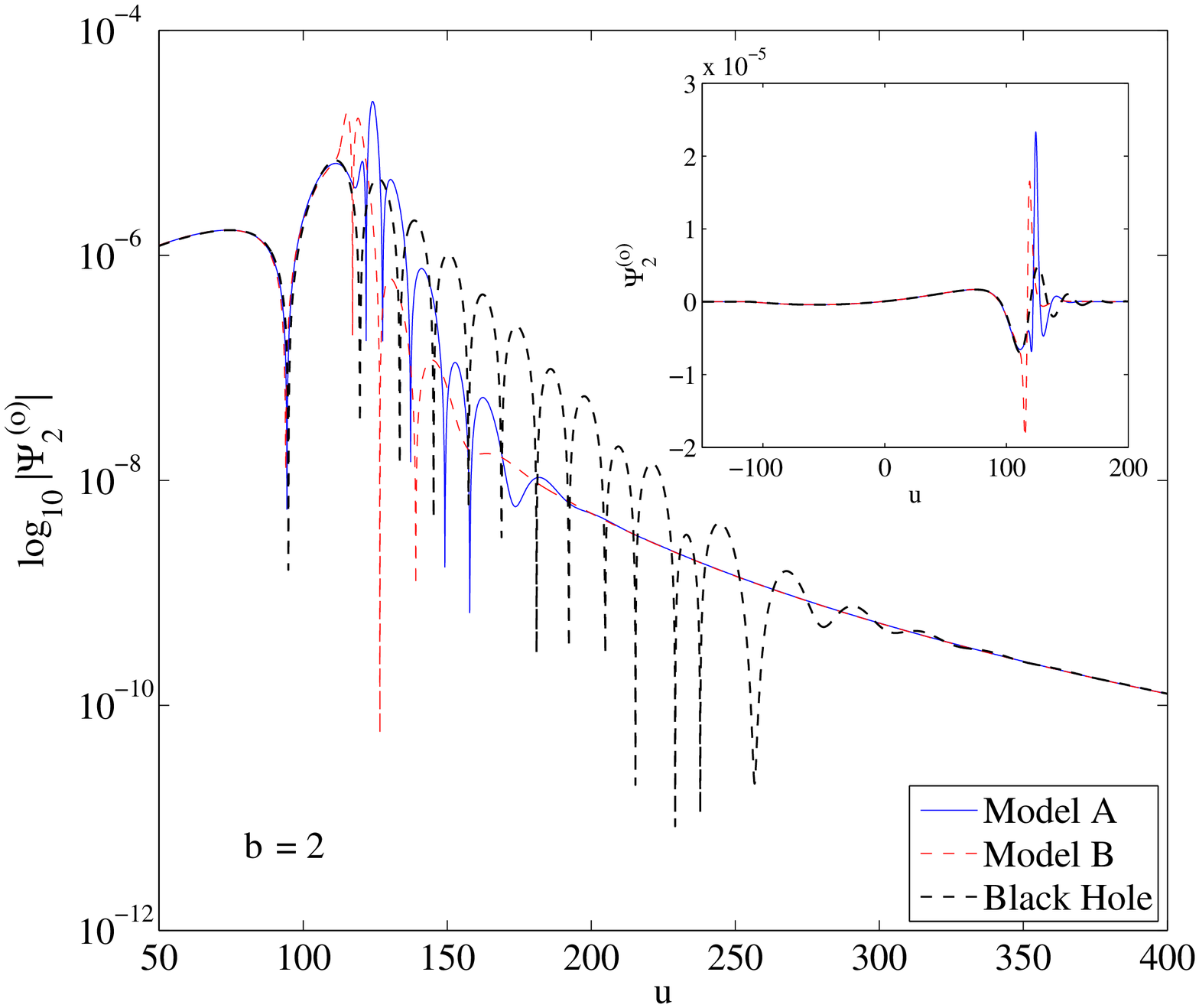}
      \vspace{-2mm}
      \includegraphics[width=85 mm]{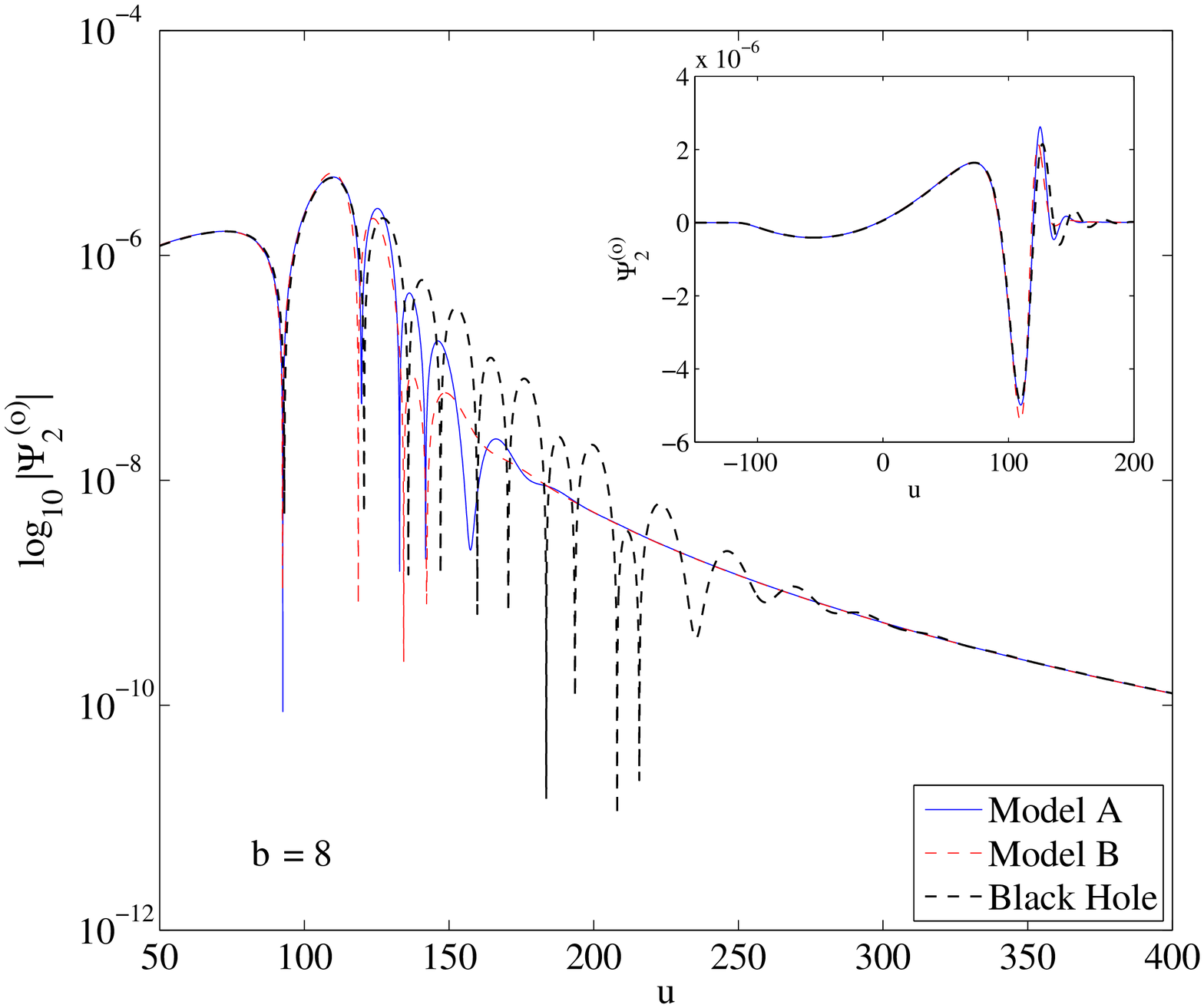}
      \vspace{-2mm}
      \includegraphics[width=85 mm]{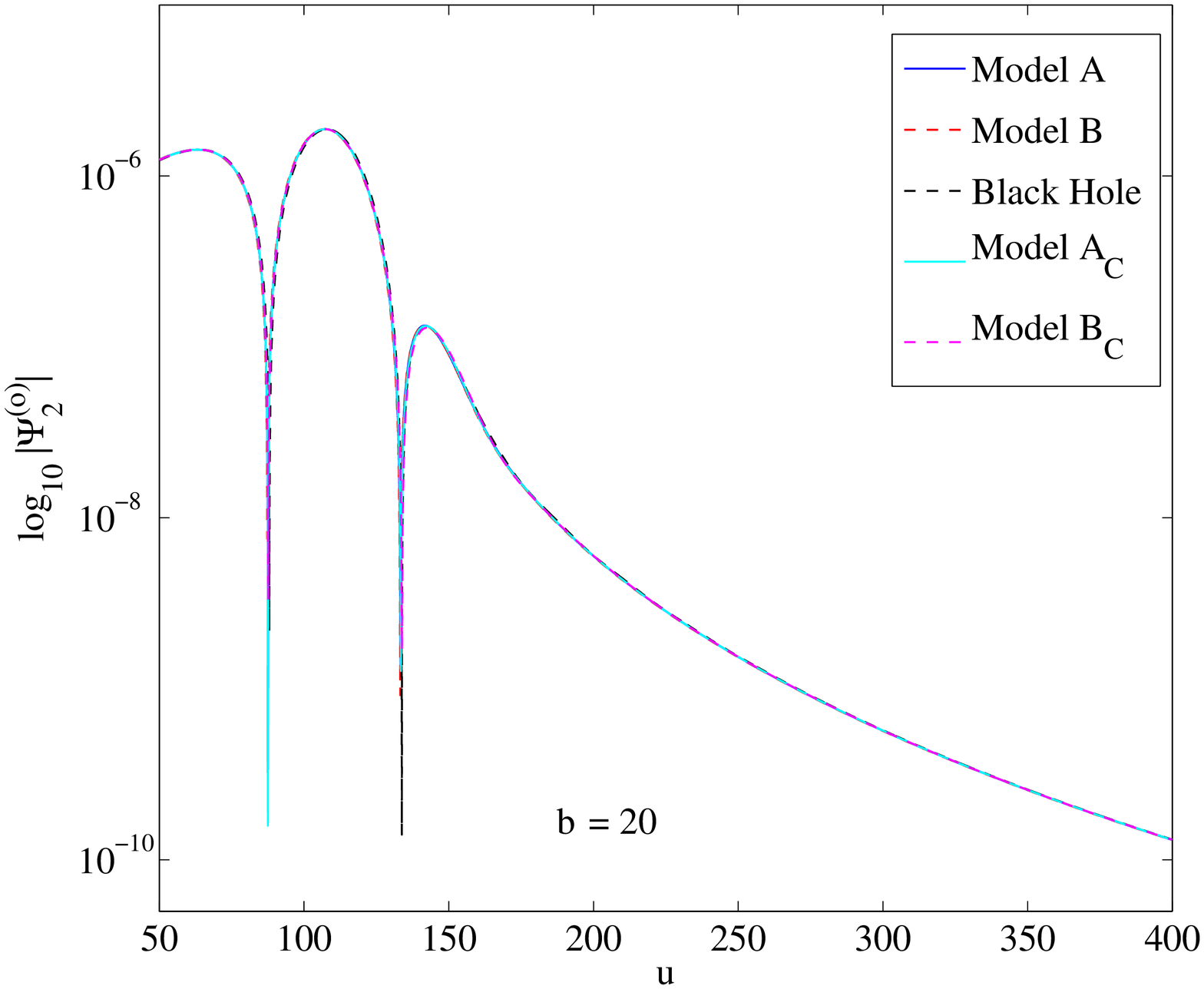}
    \end{center}
    \vspace{-4mm}
    \caption{\label{label:fig1}
      Dependence of the ring-down phase on the width 
      $b$ of the Gaussian pulse: for $b=2$ ({\it top panel}) and $b=8$
      ({\it middle panel}) the process of excitation of the space-time modes  
      shows the same qualitative features for the Black Hole and for the star. 
      The waveforms for $b=20$ ({\it bottom panel)} show there is basically no 
      difference between the gravitational wave signal backscattered from a stars
      of Table~\ref{tab:table1} and from a Schwarzschild Black Hole with the same mass.}
  \end{figure}
  
  We analyzed the gravitational wave response of relativistic stars described by
  the four (two polytropic and two constant energy density) models
  in Tab.~\ref{tab:table1} and of
  a Black Hole of the same mass to an impinging gravitational wave-packet of the
  form~\eqref{eq:ID}. We focus on the dependence of the excitation of the star
  $w$-modes (and of the Black Hole QNMs) ring-down on the width $b$.
  The Gaussian is centered at $r_0=100 $; the waveforms are extracted at 
  $r^{\rm obs}=900$ ($r_*^{\rm obs}=916$) 
  and shown versus observer retarded time $u=t-r_*^{\rm obs}$.
  Fig.~\ref{label:fig1} exhibits the waveforms, for Model~A, Model B and the black
  hole for $b=2$ (top), $b=8$ (middle) and $b=20$ (bottom). The main
  panel depicts the modulus on a logarithmic scale, in order to highlight the
  late-time non-oscillatory tail.
  
  \begin{figure}
    \begin{center}
      \includegraphics[width=80 mm, height=80 mm]{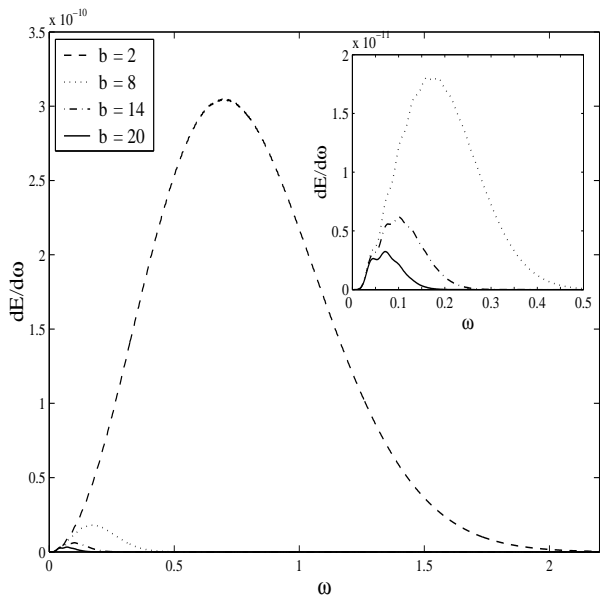} 
    \end{center}
    \vspace{-8mm}
    \caption{\label{label:fig2}Energy spectra (from Model A) for different values 
      of $b$. The maximum frequency is consistent with $\omega_{\rm max}\simeq 3\sqrt{2}/b$. 
      See text for discussion.}
  \end{figure}
  
  Let us first discuss the main features of the signal of
  Fig.~\ref{label:fig1}, starting with the ``narrow''pulse,  $b=2$.
  In the case of the Black Hole, the ring-down has the ``standard'' shape
  dominated by the fundamental mode that is quoted in textbooks. 
  In the case of the stars, a damped harmonic oscillations due to $w$-modes 
  appears (we shall make this statement more precise below). The waveforms 
  show the common {\it global} behavior {\it precursor - burst - ring-down - tail}. 
  The precursor is determined by the choice of initial data and by the long-range 
  features of the potential; this implies that, until  $ u\simeq 100$,
  the three waveforms are superposed. At later times, the short-range structure
  (burst-ring-down) becomes apparent.
  For the Black Hole the burst is related to the pulse passing through the
  peak of Regge-Wheeler potential. 
  After the pulse the quasi-harmonic oscillatory regime
  shows up. When $b$ is increased ($b=8$), the 
  features remain unchanged, but, although the non-oscillatory tail is not
  dominating yet, the amplitude of the damped oscillation is smaller and 
  lasting for a shorter time. A further enlargement of the Gaussian causes the 
  ingoing pulse to be almost completely reflected back by the  ``tail'' of the 
  potential, so that the emerging waveform is unaffected by the properties of the 
  central object. The bottom panel of Fig.~\ref{label:fig1} highlights this 
  effect for $b=20$: no quasi-normal oscillations are present. 
  It turns out that the waveforms are perfectly superposed and 
  any characteristic signature of the Black Hole or of the star 
  (for any star model, see below) disappears. 
  We have checked through a linear fit that the tail is (asymptotically) 
  in perfect agreement with the Price law: $t^{-2\ell+3}$~\cite{price1972,price1972bis}.  
  
  The absence of QNMs for large values of $b$ is qualitatively
  explained by means of the following argument 
  (see also Sec.~IX of Ref.~\cite{andersson95a}):
  in the frequency domain, the Gaussian perturbation Eq.~(\ref{eq:ID}) is 
  equivalent to a Gaussian of variance $\sigma_{\omega}=\sqrt{2}/b$ 
  and contains all frequencies. However this means that the 
  amplitudes of the modes excited by this kind of initial
  data will be exponentially suppressed if their frequencies are
  greater than the one corresponding to three standard deviations, 
  i.e., if their frequency is greater than a sort of effective maximum 
  frequency given by $\omega^{b}_{\rm max}\simeq 3\sigma_{\omega}=3\sqrt{2}/b$.
  Generally speaking, we expect to trigger the space-time modes of
  the star (or of the Black Hole) only when $b$ is such that
  $\omega^{b}_{\rm max}$ is {\it larger} than the frequency of the
  least damped quasi-normal mode of the system. In order to show how
  this argument works, let us note that we have $\omega_{\rm
    max}^{2}\simeq 2.12$, $\omega_{\rm max}^{8}\simeq 0.53$, $\omega_{\rm
    max}^{14}\simeq 0.30$ and $\omega_{\rm max}^{20}\simeq
  0.21$. Table~\ref{tab:table2} lists the first 
  six $w$-modes\footnote{These numbers have been computed by a frequency 
    domain code whose characteristics and performances are described in 
    Refs.~\cite{Pons:2001xs,Gualtieri:2001cm, Benhar:1998au,fg_private}}
  of Model~A (for $\ell=2$):
  since the lowest frequency mode has $\omega_{02}\simeq
  0.29$, it immediately follows that for $b\gtrsim 14$ the $w$-mode
  frequencies can't be found in the Fourier spectrum. This argument is
  confirmed by the analysis of the energy spectra, that are depicted in
  Fig.~\ref{label:fig2}. The frequency distribution is consistent with
  the value $\omega^b_{\rm max}\simeq 3\sqrt{2}/b$ and thus the
  $w$-modes can be excited only for $b\lesssim 14.4$.  Note that
  the different amplitudes of the spectra in Fig.~\ref{label:fig2} 
  are due to the convention used for the normalization of the 
  initial data.
  
  The same argument holds for the Black Hole. Since we have 
  $M=1.4$ the fundamental QNMs frequency is $0.2669$ (See
  Tab.~\ref{tab:table3} built from Table 1 of~\cite{kokkotas:LRR}); 
  as a result, 
  one must have $b\lesssim 15.9$ to trigger the fundamental mode 
  (that dominates the signal) although the overtones (that have lower
  frequencies) can already be present in the waveform. In any case, 
  $\omega_{\rm max}^{2}$ is smaller than the 4th overtone only and,
  due to the correspondingly large damping time, this is not expected
  to give a recognizable signature in the waveform.
  
  \begin{figure}
    \begin{center}
      \includegraphics[width=85mm]{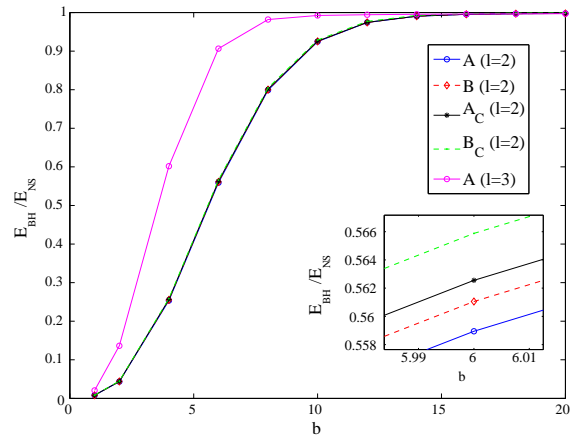} 
    \end{center}
    \vspace{-8mm}
    \caption{\label{label:fig3} The ratio, as a function of $b$, of 
      the energy released in gravitational waves by the Black Hole and
      different star models.}
  \end{figure}
  
  On the basis of these considerations, we can summarize our results 
  by saying that, for our $M=1.4M_\odot$ models, 
  when $b \gtrsim 16$, the incoming pulse is 
  totally unaffected by the short-range structure of the object and 
  the signals backscattered by any of the stars and by the Black Hole are 
  identical in practice. This information, deduced by inspecting the 
  waveforms, can be synthesized by comparing, as a function of $b$ 
  and for a fixed $\ell$, the energy released by the star ($E_{\rm NS}$) 
  and by the Black Hole ($E_{\rm BH}$) computed from Eq.~(\ref{eq:energy}). 
  Figure~\ref{label:fig3} exhibits the ratio $E_{\rm BH}/E_{\rm NS}$ for $\ell=2$
  and $\ell=3$ (the latter for Model~A only). This quantity decreases with $b$ 
  because (see Ref.~\cite{vishveshwara70}) for small $b$ the Black Hole,
  contrarily to the star, partly absorbs and partly reflects the incoming radiation.
  On the other hand, the ratio tends to one for $b\gtrsim 16$, in good numerical 
  agreement with the value of the threshold, needed to excite the quasi-normal modes, 
  that we estimated above. Notice that the saturation to one for $\ell=3$ 
  occurs for values of $b$ {\it smaller} than for $\ell=2$. This is expected: 
  in fact, the QNMs frequencies increase with $\ell$ and thus one needs 
  narrower $b$ (and thus a larger $\omega^b_{\rm max}$) to trigger space-time 
  vibrations. 
  
  \begin{table}[t]
    \caption{\label{tab:table2}The first four frequencies $\nu_{n2}$ and 
      damping times $\tau_{n2}$ of $w$-modes (for $\ell=2$) of Model~A: 
      they have been computed by means of a frequency domain code described 
      in Ref.~\cite{Pons:2001xs,Gualtieri:2001cm,Benhar:1998au}. The third
      and fourth column of the table list the corresponding {\it complex}
      frequencies $\omega_{n2}-{\rm i}\alpha_{n2}$ in our standard units. 
      We have $\omega_{n2}=2\pi\nu_{n2}M_{\odot}G/c^3$.}
    \begin{ruledtabular}
      \begin{tabular}{ccccc}
  	$n$ & $\nu_{n2}$ [Hz] & $\tau_{n2}$ [$\mu$s] 
        & $\omega_{n2}$ & $\alpha_{n2}$   \\
  	\hline
  	\hline
  	0 & 9497  & 32.64 & 0.29393  & 0.15091    \\
  	1 & 16724 & 20.65 & 0.5176   & 0.23853     \\
  	2 & 24277 & 17.21 & 0.75136  & 0.28621     \\ 
  	3 & 32245 & 15.43 & 0.99796  & 0.31923     \\
      \end{tabular}
    \end{ruledtabular}
  \end{table}
  
  \begin{table}[t]
    \caption{\label{tab:table3} The first four {\it complex}
      $\ell=2$ QNMs frequencies $\omega_{n2}-{\rm i}\alpha_{n2}$ of a $M=1.4$ 
      Black Hole in our standard units (Derived from the 
      values published  in Table 1 of~\cite{kokkotas:LRR}).}
    \begin{ruledtabular}
      \begin{tabular}{ccccc}
        $n$ & $\nu_{n2}$ [Hz] & $\tau_{n2}$ [$\mu$s] 
        & $\omega_{n2}$ & $\alpha_{n2}$ \\
  	\hline
  	\hline
  	0 & 8624 & 77.52 & 0.2669 & 0.0635 \\
  	1 & 8002 & 25.18 & 0.2477 & 0.1956 \\
  	2 & 6948 & 14.42 & 0.2150 & 0.3416 \\ 
  	3 & 5804 & 9.78  & 0.1796 & 0.5037 \\
      \end{tabular}
    \end{ruledtabular}
  \end{table}
  
  \subsection{Identification of the $w$-modes}
  
  We conclude this section by discussing the possibility of
  identifying {\it unambiguously} the presence of $w$-modes in 
  the waveforms and in the corresponding energy spectrum. Ideally, one
  would like to find precise answers to the following points:
  (i) understand  which part of the waveform can be written as 
  a superposition of $w$-modes; (ii) {\it how many} modes one should 
  expect to be excited and (iii) how does this depend on $b$. 
  
  Although these questions have been widely investigated in the 
  past (see for example Chapter~4 of~\cite{fn_98}, Ref.~\cite{ks_99} 
  and references therein), still they have not been exhaustively
  answered in the literature. The major conceptual problems underlying 
  this difficulty are (i) the fact that the quasinormal-modes sets 
  are not complete and (ii) the so called {\it time shift problem}.
  The former is intrinsic in the definition of the quasinormal modes 
  and prevents, in fact, to associate an energy to each excitation mode. 
  The latter is related to the exponential decay of the quasinormal modes 
  and it implies that, if the same signal occurs at a later time, the 
  magnitudes of the modes will be larger with respect to that of 
  the same signal occurred at an earlier time. As a consequence, 
  the use of the magnitude of the amplitudes $C_n$ 
  (see Eq.~\ref{eq:template} below) is not a good measure of the 
  excitation of the quasinormal modes. We refer to the review of 
  Nollert~\cite{nollert:1999} for a thorough discussion of such problems.
  
  \begin{figure}
    \begin{center}
      \includegraphics[width=85 mm]{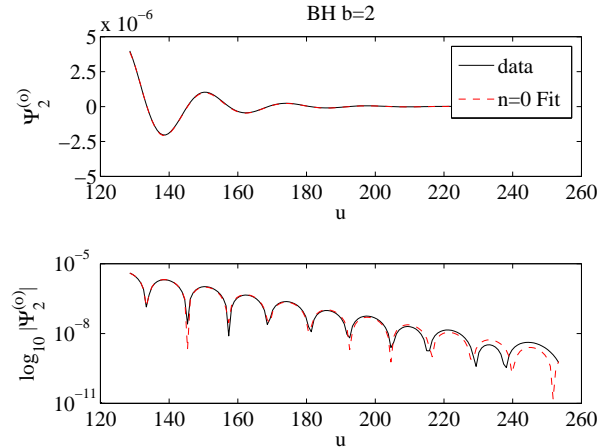}
    \end{center}
    \vspace{-5mm}
    \caption{\label{label:fitBH} Fits of the ring-down part of the waveform 
      with the fundamental ($n=0$) space-time mode for a Black Hole
      excited by a $b=2$ Gaussian pulse. We show the waveform  $\Psi^{(\rm o)}_2(t)$ and its 
      absolute value on a logarithmic scale to highlight the differences with
      the fit.
    }
  \end{figure}
  
  \begin{figure}
    \begin{center}
      \includegraphics[width=85 mm]{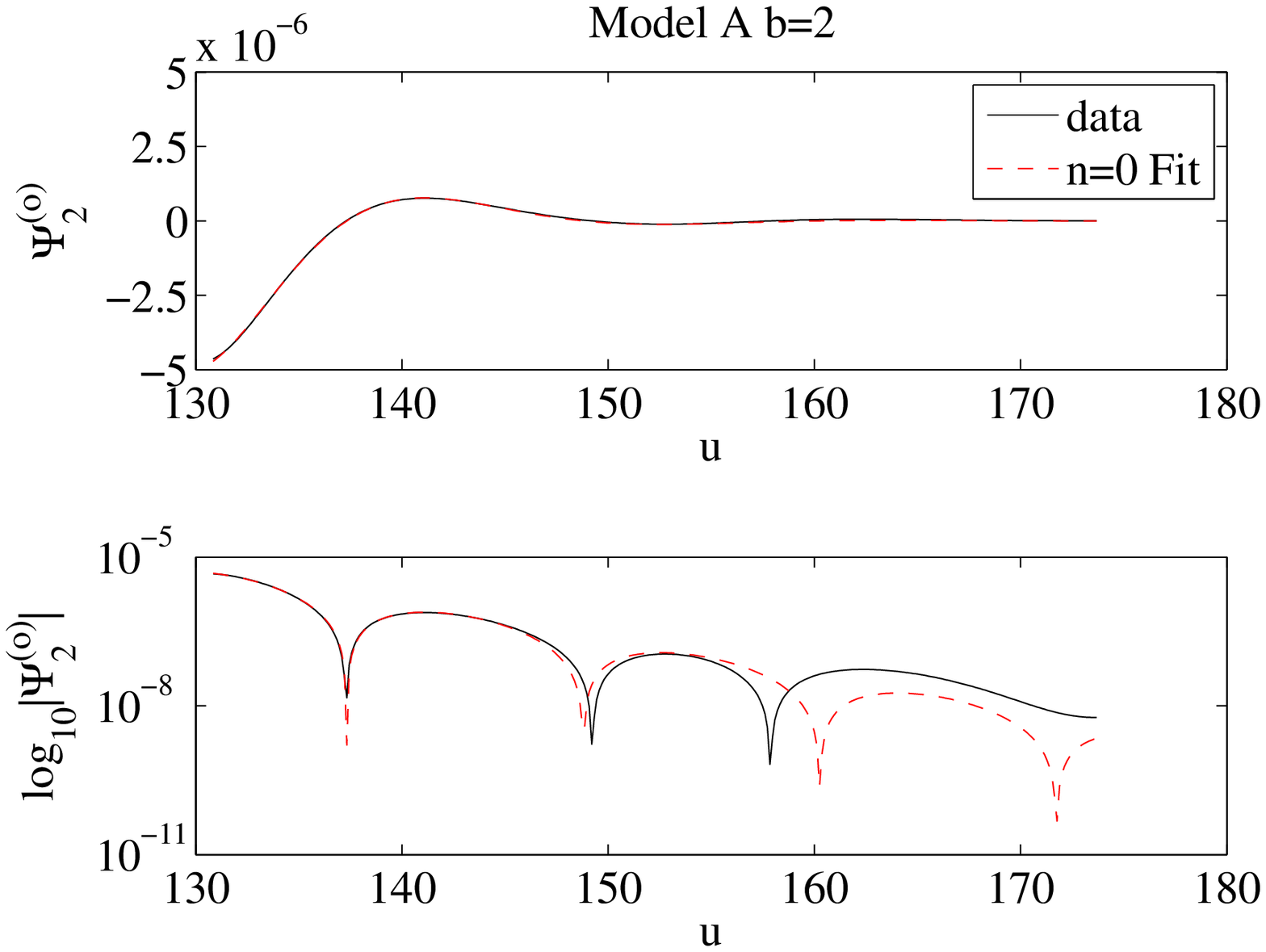}
      \includegraphics[width=85 mm]{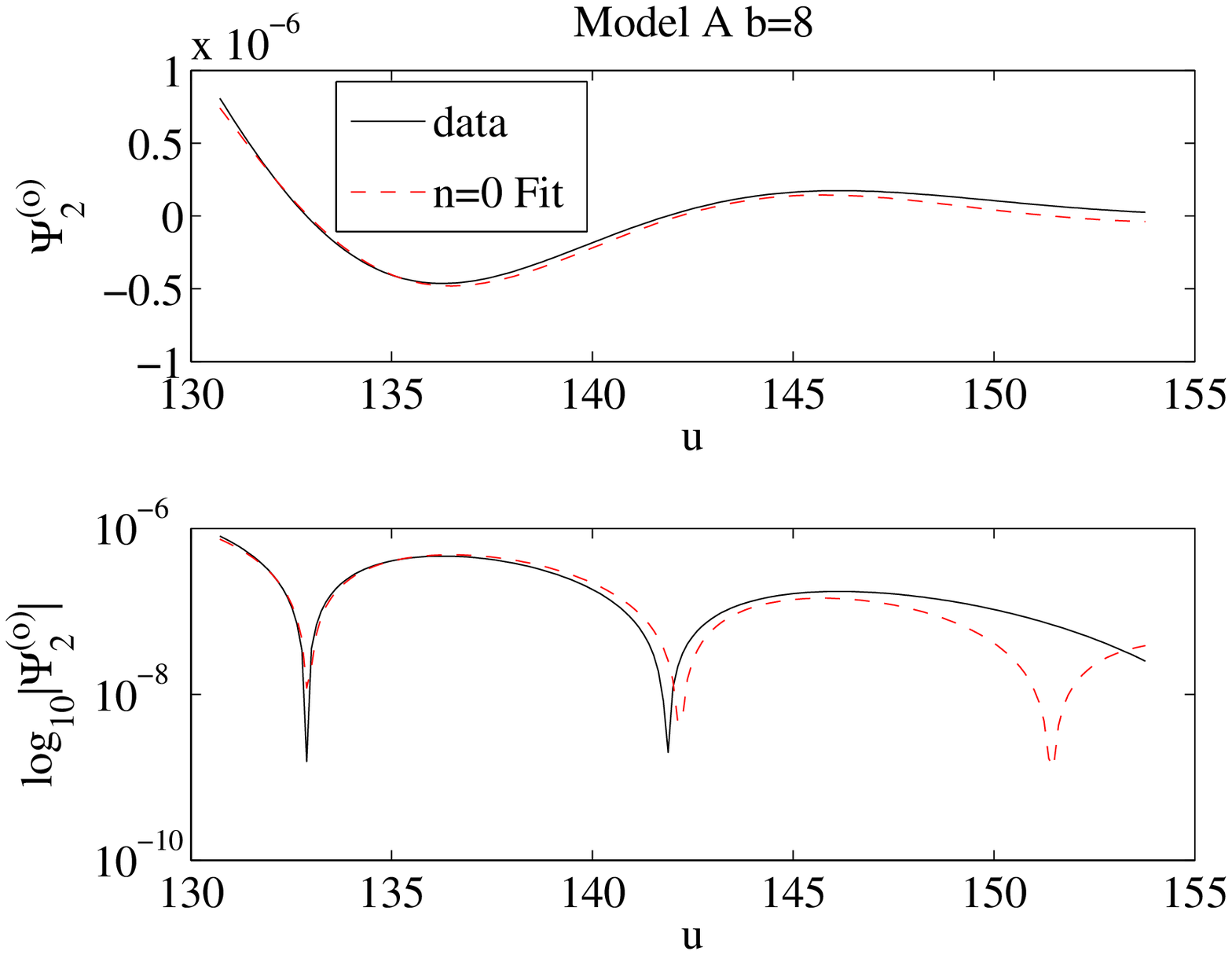}
    \end{center}
    \vspace{-5mm}
    \caption{\label{label:fitModelA} Fits of the ring-down part of the waveform 
      with the fundamental ($n=0$) space-time  mode for the stellar
      Model~A exited by $b=2$ (upper panels) and a 
      $b=8$ (bottom panels) Gaussian pulse; 
      For each value of $b$ we show the waveform  $\Psi^{(\rm o)}_2(t)$ and its 
      absolute value on a logarithmic scale to highlight the differences with
      the fit.
    }
  \end{figure}
  
  Beside these conceptual difficulties, from the practical point of view it
  is however important to extract as much as information as possible about
  the quasi-normal modes by analyzing the ringing phase of the signal.
  Two complementary methods can be used to obtain such important knowledge.
  On the one hand, one can implement the Fourier analysis, namely looking 
  at the energy Fourier spectrum in the frequency range where $w$-modes are 
  expected~(see e.g. Refs.~\cite{allen98,fk_00,passamonti06a}).
  On the other hand, one can perform a ``fit analysis''. In this case, 
  it is assumed that, on a given interval $\Delta u=[u_{\rm i},u_{\rm f}]$, 
  the waveform can be written as a superposition of $n$ exponentially 
  damped sinusoids, the quasi-normal modes expansion:
  \begin{equation}
    \label{eq:template}
    \Psi_{\ell} = \sum_{n=0}\Psi_{n\ell} = \sum_{n=0} C_{n} \cos(\omega_{n} u
    +\phi_{n})\exp(-\alpha_{n} u) \ ,
  \end{equation}
  of frequency $\omega_{n}$ and damping time $1/\alpha_{n}$, that are,
  a priori, unknown [we omit henceforth the index $\ell$ since in
    the following we will be focusing only on the $\ell=2$ modes].
  Using a non-linear fit procedure one can estimate the values
  of ($\omega_n$, $\alpha_n$, $C_{n}$ $\phi_{n}$) from the waveform.  
  We perform this analysis by means of a modified least-square Prony 
  method (see e.g. the discussion of Ref.~\cite{berti:2007}) to fit 
  the waveforms. A feedback on the reliability of our fit procedure 
  is done by comparing the values of frequency and damping time, 
  $\omega_{n\ell}$ and $\alpha_{n\ell}$, obtained by the fit with 
  those of Table~\ref{tab:table2} and Table~\ref{tab:table3} that 
  we assume to be the correct ones.
  
  The typical outcome of the fit analysis, using only the fundamental mode 
  ($n=0$), are shown in Fig.~\ref{label:fitBH} for the Black Hole 
  with $b=2$  and in Fig.~\ref{label:fitModelA}
  for the star Model~A with $b=2$ (top panel) and $b=8$ (bottom panel). 
  When $b=2$, for which the largest space-time mode excitation is expected,
  for both the star and the Black Hole the fits show excellent agreement
  with the numerical waveform at early times, that progressively worsen
  due to the power-law tail contribution. The reliability of the procedure
  is confirmed by the values of $\omega_0$ and $\alpha_0$ that we obtain
  from the fit.
  
  \begin{figure}
    \begin{center}
      \includegraphics[width=85 mm]{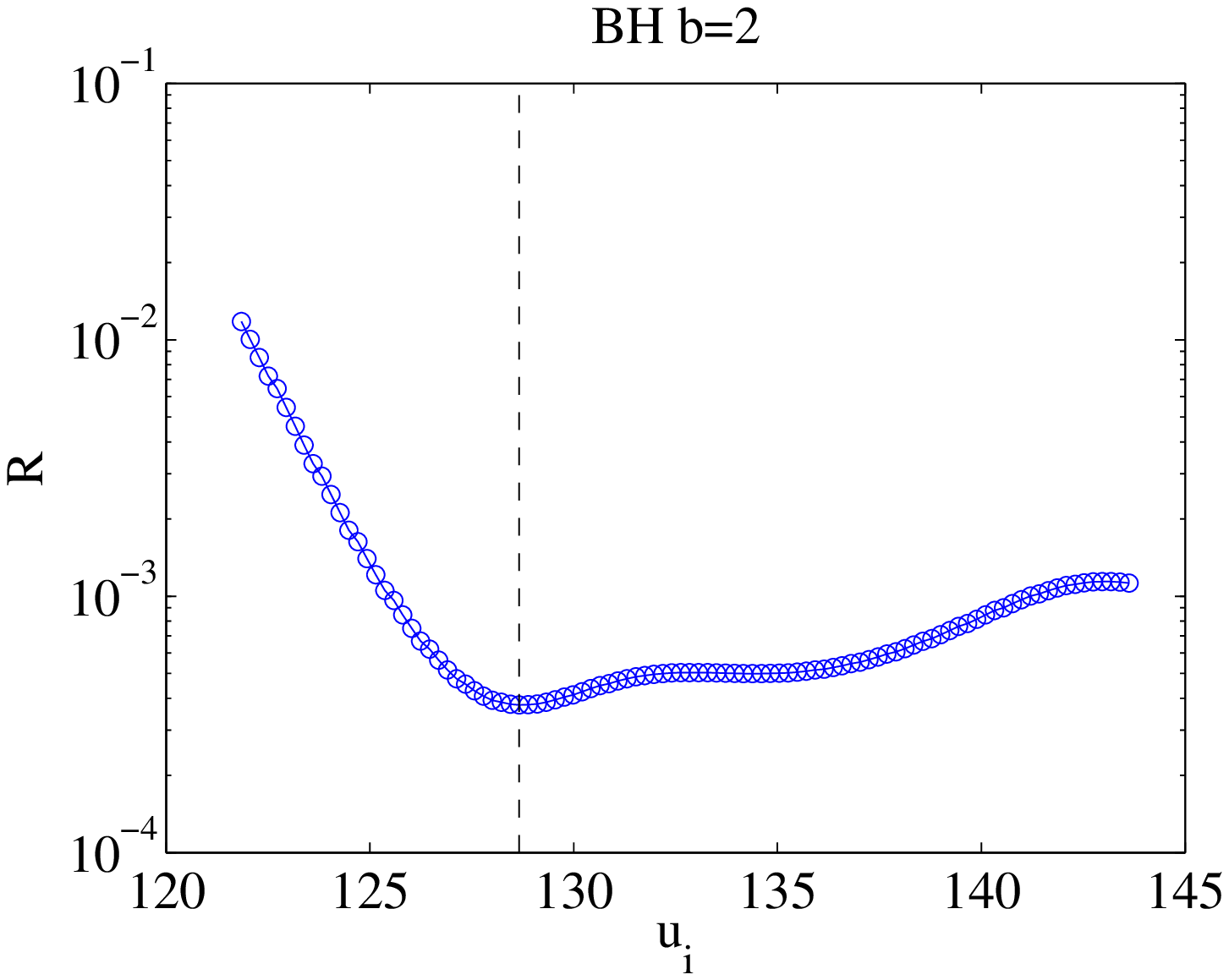}
      \includegraphics[width=85 mm]{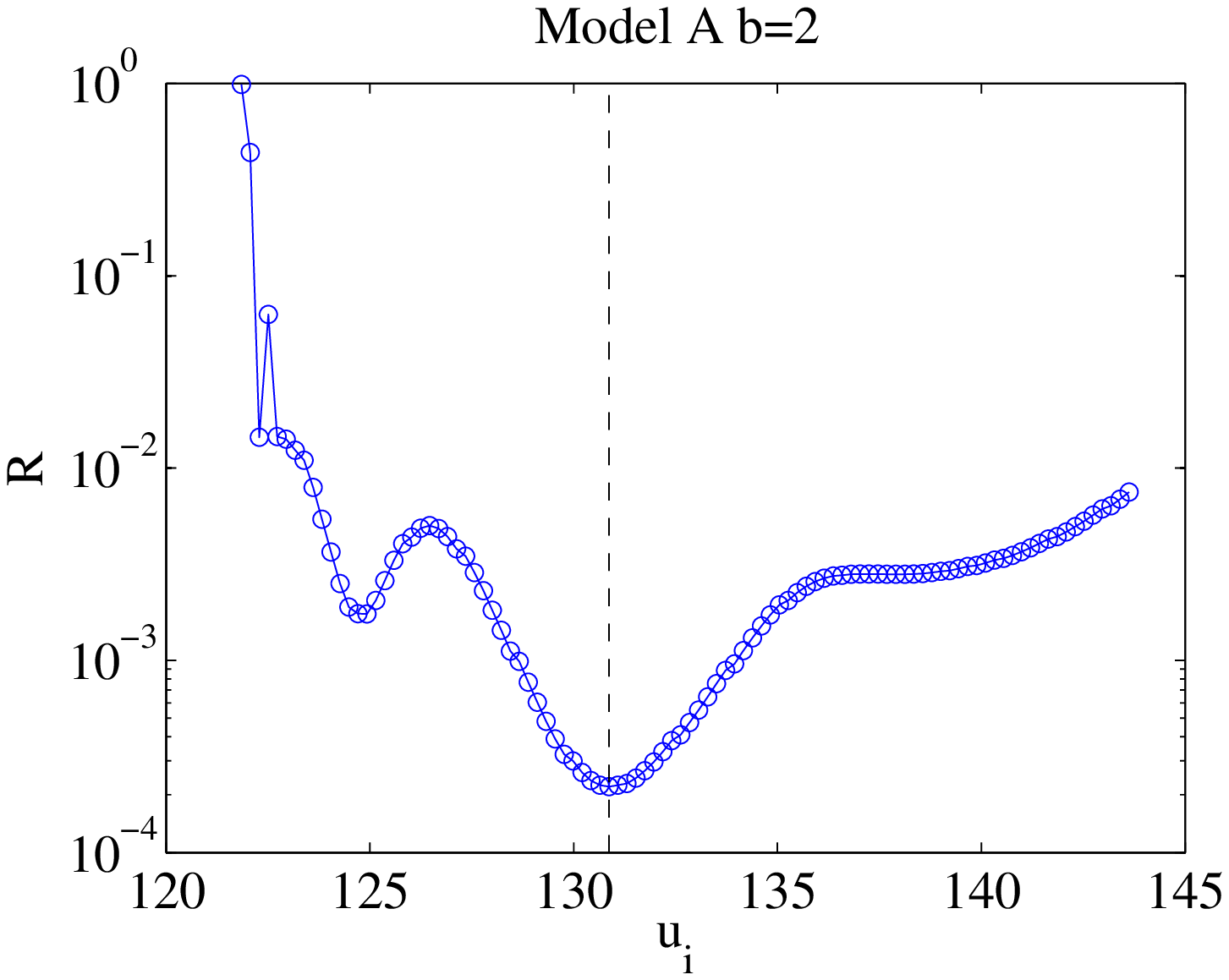}
    \end{center}
    \vspace{-5mm}
    \caption{\label{label:Residual} The residual $\mathcal{R}\equiv1\!-\!\Theta$ 
      (See Eq.~\ref{eq:residual}) 
      of the fits of the waveform 
      of the response to a $b=2$ Gaussian pulse of a Black Hole (upper panel) and stellar
      Model~A (bottom panel) as a function of the initial time ($u_{\rm i}$) of the fitting 
      window around its best values that it is $u_{\rm i}=129$ for a Black Hole and 
      $u_{\rm i}=131$ for Model~A.
    }
  \end{figure}
  
  For the Black Hole, we have $\omega_0=0.2660$ and 
  $\alpha_0=0.0631$, which differ  of respectively $0.3\%$ and $0.6\%$ 
  from the  ``exact'' values of Table~\ref{tab:table3}. We can thus 
  conclude that the fundamental mode is essentially the only mode 
  excited for $b=2$.
  For the star, Model~A, we obtain $\omega_0=0.2739$ and $\alpha_0=0.1636$ 
  and they differ respectively of $7\%$ and $8\%$ from the  ``exact'' values
  of Table~\ref{tab:table2}. Since the damping time of the fundamental star 
  $w$-mode is generically smaller than that of an equal mass Black Hole,
  the ringing is shorter and it is more difficult to obtain precise 
  quantitative statements.  In this case we tried to include more modes in the 
  template (\ref{eq:template}) used for the fit in order to precisely 
  quantify the real contribution due to the presence of
  overtones in the signal. Unfortunately, in this case the fit 
  procedure seems badly conditioned and we could not obtain 
  a sensible feedback of the frequencies even if we clearly obtain 
  (having more adjustable parameters) a better fit. 
  
  \begin{figure*}[t]
    \begin{center}
      \includegraphics[width=85 mm]{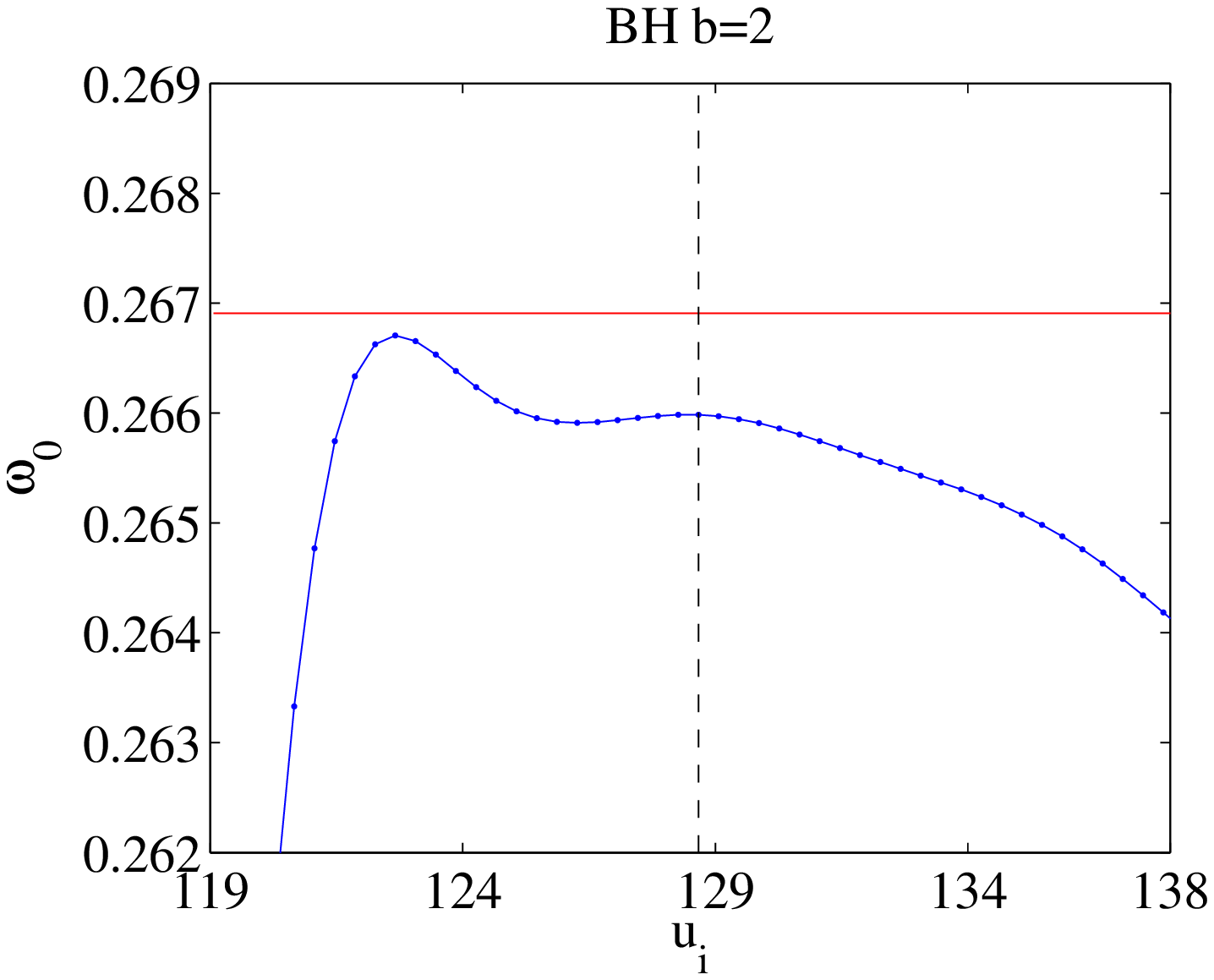}
      \includegraphics[width=85 mm]{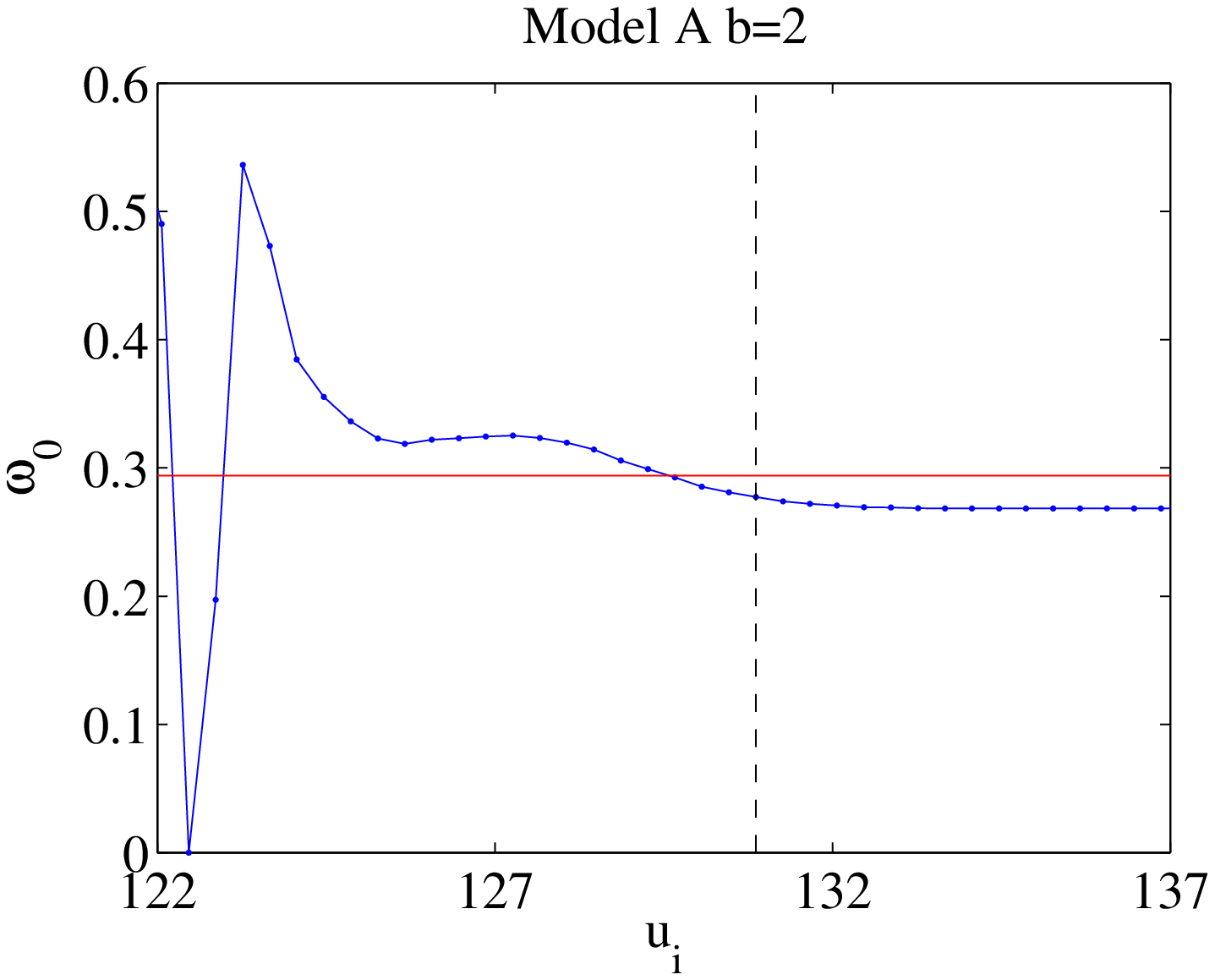}\\[2mm]
      \includegraphics[width=85 mm]{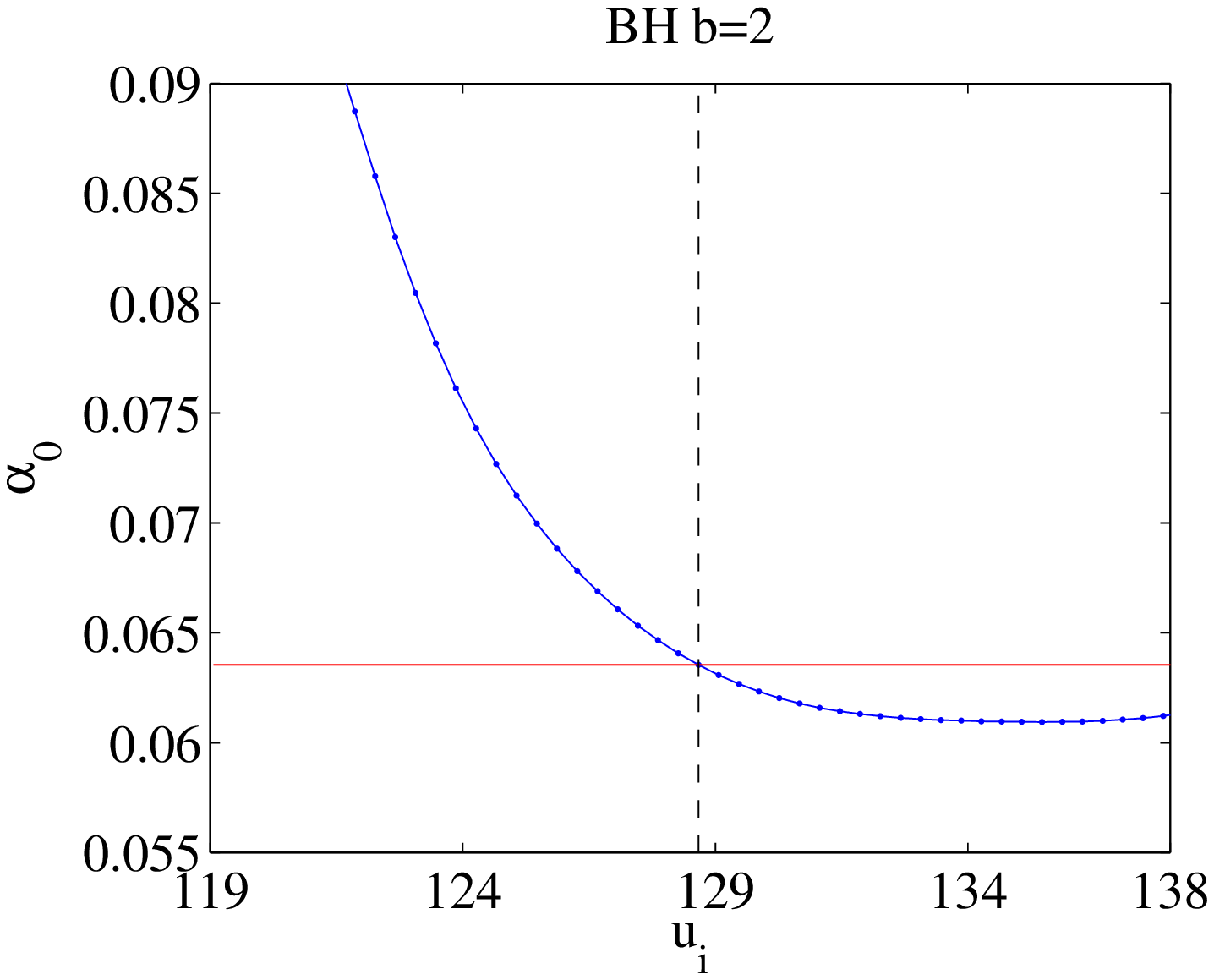}
      \includegraphics[width=85 mm]{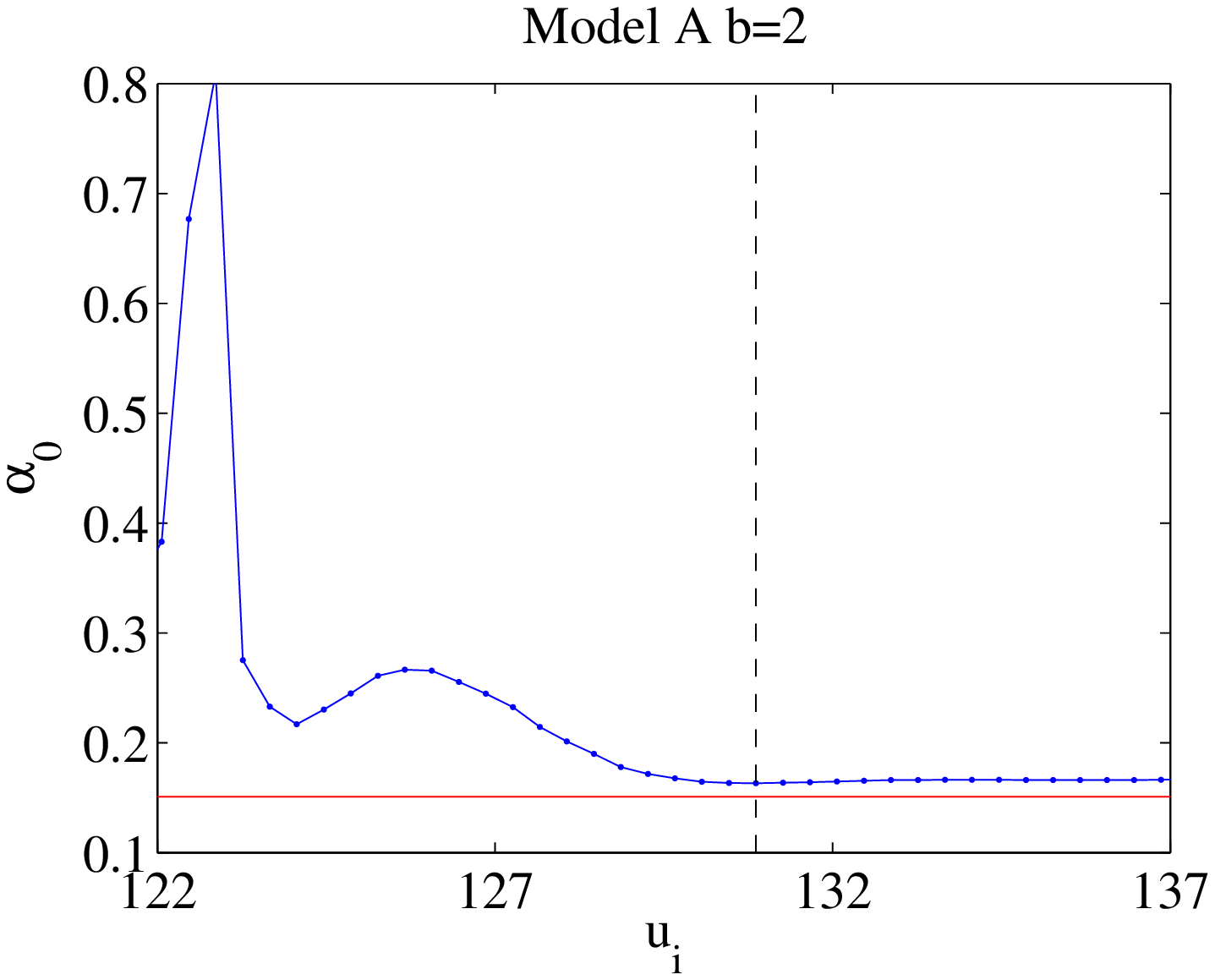}
    \end{center}
    \vspace{-5mm}
    \caption{\label{label:FitModels} Determination of the best window for the fit
      of the Black Hole waveform and Model~A ($b=2$). The initial time $u_{\rm i}$ 
      is chosen so to  minimize the residual $\mathcal{R}$ between the 
      data and the fit. For the Black Hole we obtain $u_{\rm i}=129$ ($u_{\rm f}=254$), 
      while for Model~A we have  $u_{\rm i}=131$ ($u_{\rm f}=174$). 
      The horizontal line indicates the ``exact'' values of the considered model
      reported in Table~\ref{tab:table2} and Table~\ref{tab:table3}.  
    }
  \end{figure*}
  
  We have found that the choice of the time window to perform this analysis 
  has a strong influence on the result of the fit. This choice is delicate 
  and it is related to the aforementioned problem of the time shift. Ideally, 
  the window should start with the ring-down (i.e., at the end of the {\it burst}) 
  and it must be both sufficiently narrow, in order not to be influenced by 
  the non-oscillatory tail, and sufficiently extended to include all the relevant 
  information. There are no theoretical ways to predict or estimate the correct window, 
  but some systematic procedures have actually been explored in the literature.
  We decided to use a method very similar to the one discussed in details in 
  Ref.~\cite{Dorband:2006gg}: it consists in setting $u_{\rm f}$ at the end 
  of the oscillatory phase, which is clearly identifiable in a logarithmic plot, 
  and choosing the initial time of the window $u_{\rm i}$ such as to minimize 
  the difference between the real data ($\Psi^{\rm data}_j$) and the 
  waveform synthesized from the results of the fitting procedure
  ($\Psi^{\rm fit}_j$). This difference is estimated 
  by means of the following ``scalar product''
  \begin{equation}
    \label{eq:residual}
    \Theta(\Psi^{\rm data},\Psi^{\rm fit}) 
    \equiv \frac{\sum_j \Psi^{\rm data}_j \Psi^{\rm fit}_j}
  	   {\sqrt{\sum_j (\Psi^{\rm data}_j)^2}\sqrt{\sum_j (\Psi^{\rm fit}_j)^2}} 
  \end{equation}
  whose result $\Theta$ is a value in the interval $[0,1]$ that it is exactly one
  when the two time series are identical (perfect fit).
  Fig.~\ref{label:Residual} shows such a determination for the Black Hole (top panel) 
  and Model~A (bottom panel) with $b=2$: both curves exhibit a clear minimum 
  of the quantity $\mathcal{R}\equiv1\!-\!\Theta$ at, respectively, $u_{\rm i}=129$ and  
  $u_{\rm i}=131$. The time window extends to $u_{\rm f}=254$ (Black Hole) 
  and $u_{\rm f}=174$ (Model~A), respectively.
  
  As can be seen in Fig.~\ref{label:FitModels} one has that
  even a small change of initial time $u_{\rm i}$ of the window used
  produces sensible variation of the estimated values of the frequency and the 
  damping time of the fundamental mode. However, it should be noticed that 
  the estimated values obtained for the best window are those in best agreement 
  with the expected values. 
  
  We finally repeat the analysis for the datasets relative to wider Gaussian 
  pulses. Focusing on the representative $b=8$ case, we find essentially the 
  same picture with two main differences. First, the ``global'' quality of the fit,
  given by $\mathcal{R}$ is less good than in the $b=2$ case;  this is particularly 
  evident for the star, where the fitted frequencies differ from the ``exact'' ones
  by more than $10\%$. Second, the fitting window becomes narrower and narrower 
  as $b$ is increased (see column seven of Table~\ref{tab:table4}), thereby the
  the fit analysis quickly becomes meaningless.
  
  \begin{table}[t]
    \caption{\label{tab:table4} The results of the fit of the fundamental 
      mode for the Black Hole and for the stellar Model~A in the response 
      to Gaussian pulse with $b=2$ and $b=8$ for the best fit 
      windows $[u_{\rm i},u_{\rm f}]$ determined using the minimum of the residual 
      $\mathcal{R}\equiv1\!-\!\Theta$ criteria (see Fig.\ \ref{label:Residual}). The reported error refer
      to relative difference between the fitted values and reference values
      reported in Table~\ref{tab:table2} and Table~\ref{tab:table3}.}
    \begin{ruledtabular}
      \begin{tabular}{cccccccc}
  	Model & $b$ & $\omega_0$ & $\delta\omega_0$ $\%$ & $\alpha_0$
        & $\delta\alpha_0$ $\%$ & $[u_{\rm i},u_{\rm f}]$ &
        $\mathcal{R}$ \\ \hline A & 2 & 0.2739 & 7 & 0.1636 & 8 &
        [131,174] & $2\times10^{-4}$\\ A & 8 & 0.3396 & 15 & 0.1302 & 13
        & [131,154] & $1\times10^{-2}$\\ BH & 2 & 0.2660 & 0.3 & 0.0631
        & 0.6 & [129,254] & $4\times10^{-4}$\\ BH & 8 & 0.2614 & 0.2 &
        0.0591 & 0.7 & [129,229] & $5\times10^{q-3}$\\
      \end{tabular}
    \end{ruledtabular}
  \end{table}
  
  For the reasons outlined above we have found this procedure not as
  effective as we hoped. We think that the problem of the unambiguous
  determination of the ``right'' time interval for the fit and of the
  presence and quantification the overtones in numerical data deserves
  further considerations.
  
  \section{Conclusions}
  \label{sec:conclusions}
  
  In this paper we have studied numerically the scattering of odd-parity
  Gaussian pulses of gravitational radiation off relativistic stars and
  Black Holes. We have found that the excitation of $w$-modes and black
  hole QNMs occurs basically in the same way for both objects: pulses of
  small $b$ (high frequencies) can trigger the $w$-modes, while for
  large $b$ (low frequencies) one can only find curvature backscattering
  effects and non-oscillatory tails. When $w$-modes are present, we have
  shown that both frequency-domain (energy spectrum) and time-domain
  (fit to a superposition of $w$-modes) analysis are useful to
  understand the mode content of the waveforms; however, our study also
  indicates that it is difficult to single out {\it precisely} the
  contribution of each mode, since the fundamental mode always dominates
  the signal and a clear identification of the overtones is lacking in
  the case of excitation induced by the scattering of odd-parity
  Gaussian pulses of gravitational radiation.
  
  The inspiring idea of this paper was to understand the origin of the
  dynamical excitation of $w$-modes in a simple, but rather general,
  setting, where it is possible to do many, quick and controllable
  high-accuracy numerical simulations with a tunable ``source''.  Our
  expectation is that the main features of the process of $w$-mode
  excitation (i.e., its dependence on the intrinsic frequency content of
  the initial data) that we have highlighted in the perturbative regime
  are sufficiently robust to survive {\it qualitatively} also in the
  full 3D-nonlinear case that needs the solution of the full set of
  Einstein equations. An encouraging motivation for hoping so is that in
  other physical settings, like for example the merger of two equal mass
  Black Holes, the perturbative analysis of the early 70s~\cite{DRT} in
  the extreme mass ratio limit was able to single out the important
  physical elements (i.e., the presence of QNMs), sketching a picture
  that has been later refined, but substantially confirmed, by numerical
  relativity
  simulations~\cite{Pretorius:2005gq,Campanelli:2005dd,Baker:2005vv,Baker:2007fb,Pretorius:2007nq}.
  
  \section*{Acknowledgments}
  
  Computation performed on the {\tt Albert} Beowulf clusters at the
  University of Parma. We gratefully thank V.~Ferrari and L.~Gualtieri
  for computing for us the $w$-mode frequencies of
  Table~\ref{tab:table2}. The commercial software Matlab$^{\rm TM}$ has
  been used for most of the analysis reported here.  Software
  implementing the Prony methods is freely available at {\tt
    http://www.statsci.org/other/prony.html}. SB has been supported by a
  Marie-Curie fellowship during the {\it General Relativity Trimester on
    Gravitational Waves, Astrophysics and Cosmology} at Institut Henri
  Poincar\'e (Paris), where part of this work was done.  The activity of
  AN at IHES is supported by an INFN fellowship. AN also gratefully
  acknowledges support of ILIAS.

\end{document}